\documentclass[12pt,preprint]{aastex}
\singlespace
\slugcomment{Accepted for the publication in the {\it Astrophysical Journal}}
\def\farcm{\hbox{$.\mkern-4mu^\prime$}}

\def\la{\mathrel{\hbox{\rlap{\hbox{\lower4pt\hbox{$\sim$}}}\hbox{$<$}}}}
\def\ga{\mathrel{\hbox{\rlap{\hbox{\lower4pt\hbox{$\sim$}}}\hbox{$>$}}}}

\shortauthors{Park}
\shorttitle{Sgr A East}

\begin{document}
\title{A Candidate Neutron Star Associated with Galactic Center Supernova Remnant 
Sagittarius A East}
\author{Sangwook Park\altaffilmark{1,5}, Michael P. Muno\altaffilmark{2},
Frederick K. Baganoff\altaffilmark{3}, Yoshitomo Maeda\altaffilmark{4}, 
Mark Morris\altaffilmark{2}, George Chartas\altaffilmark{1}, 
Divas Sanwal\altaffilmark{1}, David N. Burrows\altaffilmark{1}, and 
Gordon P. Garmire\altaffilmark{1}}

\altaffiltext{1}{Department of Astronomy and Astrophysics, Pennsylvania 
State University, 525 Davey Laboratory, University Park, PA. 16802}
\altaffiltext{2}{Department of Physics and Astronomy, University of
California Los Angeles, Los Angeles, CA. 90095}
\altaffiltext{3}{Center for Space Research, Massachusetts Institute of
Technology, Cambridge, MA. 02139 }
\altaffiltext{4}{Institute of Space and Astronautical Science, 3-1-1 
Yoshinodai, Sagamihara, Kanagawa, 229-8510, Japan}
\altaffiltext{5}{park@astro.psu.edu}

\begin{abstract}
We present imaging and spectral studies of the supernova remnant (SNR)
Sagittarius (Sgr) A East from deep observations with the {\it Chandra X-Ray
Observatory}. The spatially-resolved spectral analysis of Sgr A East 
reveals the presence of a two-temperature thermal plasma ($kT$ $\sim$ 1 keV
and 5 keV) near the center of the SNR. The central region is dominated by 
emission from highly-ionized Fe-rich ejecta. We estimate a conservative
upper limit on the total Fe ejecta mass of the SNR, M$_{Fe}$ $<$ 0.27 M$_{\odot}$. 
Comparisons with standard SN nucleosynthesis models suggest that this Fe mass 
limit is consistent with a Type II SN explosion for the origin of Sgr A East. 
On the other hand, the soft X-ray emission extending toward the north of the SNR
can be described by a single-temperature ($kT$ $\sim$ 1.3 keV) thermal plasma 
with normal chemical composition. This portion of the SNR is thus X-ray emission 
from the heated interstellar medium rather than the metal-rich stellar ejecta. 
We point out that a hard pointlike source CXOGC J174545.5$-$285829 (the so-called 
``cannonball'') at the northern edge of the SNR shows unusual X-ray characteristics 
among other Galactic center sources. The morphological, spectral, and temporal 
characteristics of this source suggest an identification as a high-velocity  
neutron star. Based on the suggested Type II origin for the SNR Sgr A East
and the proximity between the two, we propose that CXOGC J174545.5$-$285829
is a high-velocity neutron star candidate, born from the core-collapse SN which
also created the SNR Sgr A East. 
 
\end{abstract}


\section {\label {sec:intro} INTRODUCTION}

Sagittarius (Sgr) A East, a part of the Sgr A complex in the Galactic center, 
is an extended radio source with an angular size of  2$\farcm$7 
$\times$ 3$\farcm$6 \citep{ekers83}. The nonthermal radio spectrum 
(radio spectral index $\alpha$ = 0.76, where S$\propto$$\nu$$^{-\alpha}$)
and the shell-like morphology suggested an identification of a supernova 
remnant (SNR) for Sgr A East \citep{jones74,ekers75,ekers83}. The 
{\it ROSAT}, {\it ASCA}, and {\it Beppo-SAX} satellites detected diffuse X-ray 
emission from Sgr A East, but extensive analysis was infeasible 
because of the limited detector capabilities \citep{pred94,koyama96,sidoli99}.

Recently, observations with the high angular resolution instruments aboard the 
{\it Chandra X-Ray Observatory} clearly resolved the diffuse X-ray emission 
interior to the Sgr A East radio shell (Maeda et al. 2002, M02 hereafter). 
The {\it Chandra} data revealed a centrally-peaked X-ray morphology of 
Sgr A East. The X-ray emission originates from a hot thermal plasma with 
an electron temperature $kT$ $\sim$ 2 keV, showing strong He$\alpha$ lines 
from highly-ionized elemental species of S, Ar, Ca, and Fe. The estimated low 
mass of the hot gas ($\sim$2 M$_{\odot}$) and the total thermal energy ($E_{th}$ 
= 2 $\times$ 10$^{49}$ ergs) indicated a single SNR origin for Sgr A East (M02). 
The {\it Chandra} data revealed that the Fe-rich ejecta is concentrated near 
the center of the SNR, while other elemental species are uniformly distributed 
over the SNR. M02 further suggested that Sgr A East is a metal-rich, 
mixed-morphology (MM) SNR from a Type II SN explosion of a 13$-$20 M$_{\odot}$ 
progenitor.

Sgr A East was also observed with the {\it XMM-Newton Observatory} (Sakano et al. 
2004, S04 hereafter). The results from the {\it XMM-Newton} data analysis were 
also consistent with a single SNR interpretation for Sgr A East. S04, however, 
found that a two-temperature thermal plasma ($kT$ $\sim$ 1 keV and 4 keV) was 
required in order to adequately describe the observed X-ray emission line features. 
The measured metal abundance patterns and the estimated Fe ejecta mass suggested 
either Type Ia or Type II origin. A weak 6.4 keV ``neutral'' Fe line feature was 
also detected with the {\it XMM-Newton} data. This neutral Fe line was attributed 
to the irradiation by Sgr A East of the molecular clouds which are also interacting 
with Sgr A East (S04). The discrepancies between the {\it Chandra} and the 
{\it XMM-Newton} data might have been caused by the low photon statistics of the 
{\it Chandra} data (S04). Effects of the relatively poor angular resolution of the 
{\it XMM-Newton} data may not be ruled out, either.

Since the early {\it Chandra} observations of Sgr A East presented by M02, we 
have performed follow-up observations of the Galactic center with {\it 
Chandra} as parts of the monitoring program of the Galactic center supermassive 
black hole candidate Sgr A* \citep{baganoff03}. As of 2002 June, the total 
exposure of {\it Chandra} observations of the Galactic center reached $\sim$600 
ks, representing the deepest X-ray observations of the Galactic center. Utilizing 
the wealth of these deep observations, we have presented the complex nature of 
the pointlike and diffuse X-ray emission features in the Galactic center in a 
series of publications \citep{baganoff03,muno03a,muno03c,park04a,muno04a,muno04b}. 
The current deep {\it Chandra} data increase the photon statistics of Sgr A East 
by an order of magnitude, compared with that of M02. Taking advantage of the deep 
{\it Chandra}  exposures, in this paper we present spatially-resolved spectral
analysis of Sgr A East, which was infeasible with previous data.
This approach is effective for the analysis of the ejecta material, excluding 
substantial contamination from the swept-up ISM, as successfully demonstrated with 
other Galactic SNR studies (e.g., Park et al. 2004b). The large-scale, {\it global}
spectral analysis of Sgr A East has been performed with the previous {\it Chandra} 
and {\it XMM-Newton} observations, which revealed overall spectral parameters and 
thermal characteristics of the SNR (M02; S04). Rather than presenting such general 
characteristics, we concentrate in this work on the origin of Sgr A East 
in the context of the SN explosion types. Particularly, we draw an attention to 
a hard {\it pointlike} source detected in the northern edge of Sgr A East. 
The overall X-ray characteristics suggest that this source is a candidate, 
high-velocity neutron star (NS) associated with the SNR Sgr A East. The observations 
are briefly described in \S\ref{sec:obs}. The X-ray image and spectral analyses 
of the SNR Sgr A East and the candidate NS are presented in \S\ref{sec:snr} and 
\S\ref{sec:ns}, respectively. We discuss key characteristics of the SNR and the 
NS candidate in \S\ref{sec:disc}.  Finally, a summary is presented in 
\S\ref{sec:summary}.

\section{\label{sec:obs} OBSERVATIONS \& DATA REDUCTION}

Since 1999, the Galactic supermassive black hole candidate Sgr A* has been 
monitored with the Advanced CCD Imaging Spectrometer (ACIS) \citep{garmire03} on 
board the {\it Chandra X-Ray Observatory} (Table \ref{tbl:tab1}). As of 2002 June, 
combining 11 {\it Chandra}/ACIS observations (except for ObsID 1561a, which was 
severely contaminated by a bright transient source within the field of view [FOV]; 
Muno et al. 2003b), the total exposure has reached $\sim$590 ks, which is the 
deepest ever observation of the Galactic center region in X-rays. We utilized 
the data reduced by Muno et al. (2003a), as we briefly describe here. We first 
applied the algorithm developed by Townsley et al. (2002a) to correct the spatial 
and spectral degradation of the ACIS data caused by radiation damage, known 
as charge transfer inefficiency (CTI; Townsley et al. 2000). 
We then screened the data by status, grade, and the flight timeline filter. 
We have also removed observation time intervals of strong flaring in the 
background. All individual event files were then reprojected to the tangent 
plane at the radio position of Sgr A* (RA[J2000] = 17$^h$ 45$^m$ 40$^s$.0409, 
Dec[J2000] = $-$29$^{\circ}$ 00$'$ 28$\farcs$118) in order to generate the 
composite data. The detailed descriptions of the data reduction process
and the resulting broadband raw image from the composite data may be found 
in Muno et al. (2003a).

\section{\label{sec:snr} Sgr A East}

\subsection{\label{subsec:snrimage} X-Ray Images}

\begin{figure}[]
\figurenum{1}
\centerline{\includegraphics[angle=0,width=0.5\textwidth]{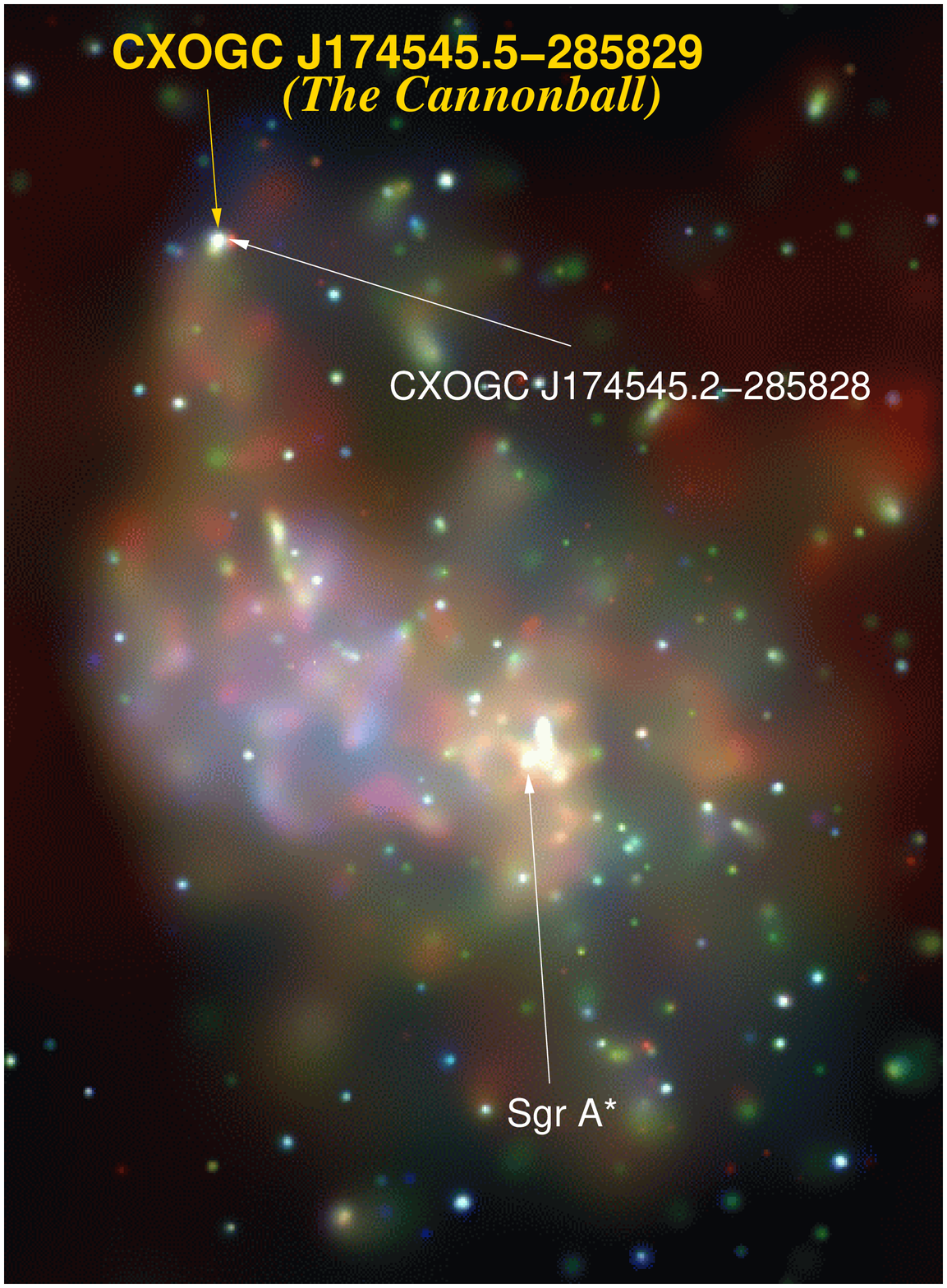}
\includegraphics[angle=0,width=0.55\textwidth]{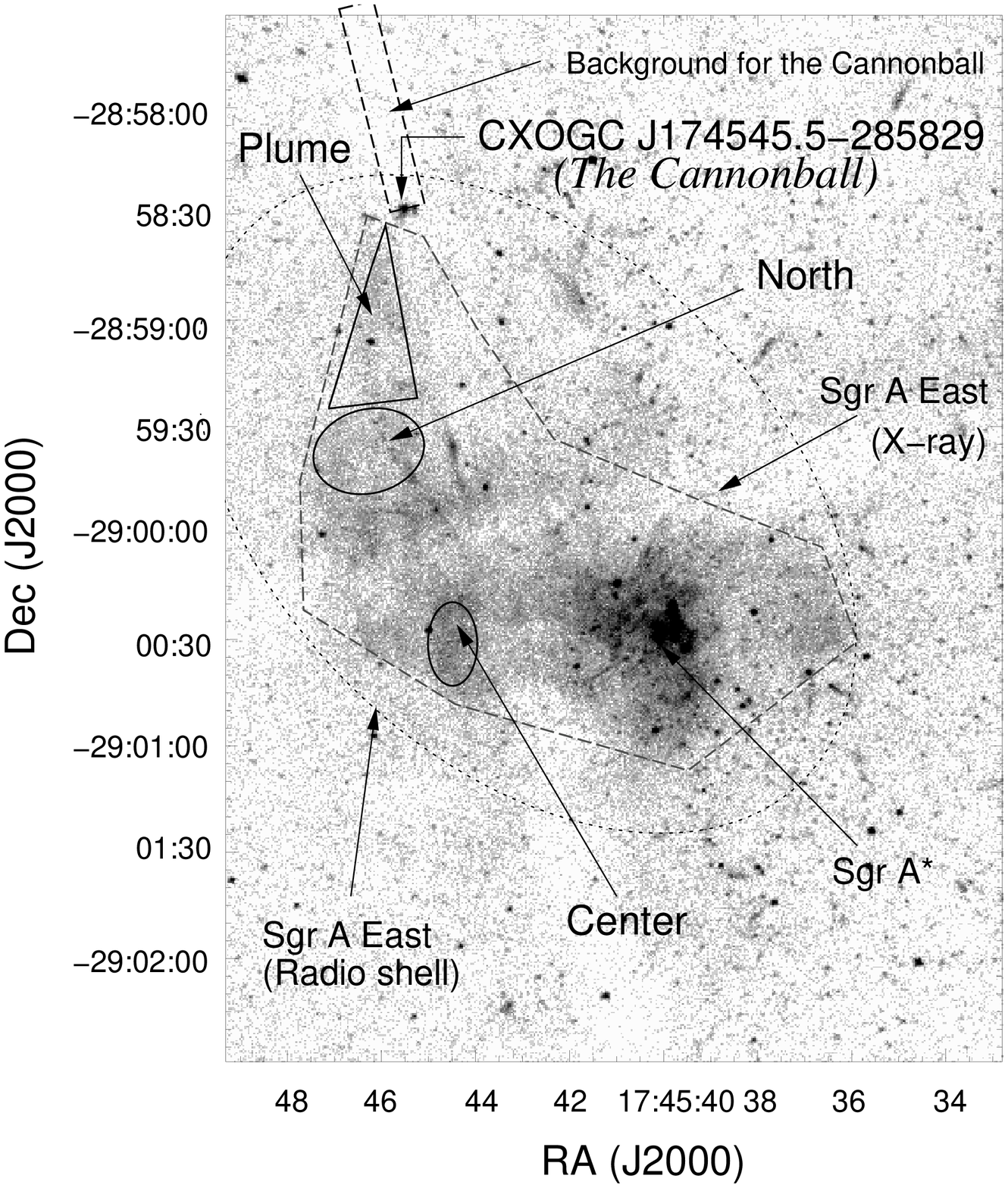}
}
\figcaption[]{{\it Left:} The exposure-corrected true color image of Sgr A East
from the composite data of 11 {\it Chandra}/ACIS observations: red represents 
the 1.5 $-$ 4.5 keV band photons, green is 4.5 $-$ 6.0 keV, and blue is 6.0 $-$ 
8.0 keV band. Each subband image has been adaptively smoothed and the detected 
point sources have not been removed. {\it Right:} The gray-scale, broadband 
(1.5 $-$ 8.0 keV) ACIS image of Sgr A East. Darker gray scales are higher
intensities. This broadband image is unsmoothed, and has not been 
exposure-corrected. The ACIS-I chip-gaps are evident: i.e., the ``whiteish (or
low intensity) lanes'' bisecting the SNR by north-south and east-west. Some key 
features within the FOV and regions used for the spectral analysis are marked.
\label{fig:fig1}}
\end{figure}

The center of Sgr A East radio shell is $\sim$50$^{\prime\prime}$ northeast of Sgr A* 
and thus the entire SNR Sgr A East was imaged within the 17$'$ $\times$ 17$'$ FOV 
of the ACIS-I array during the Sgr A* monitoring observations. A ``true-color'' 
X-ray image of the Sgr A East is presented in Figure~\ref{fig:fig1}. Each subband 
image has been exposure-corrected utilizing the exposure map produced by Muno et al. 
(2003a), and adaptively smoothed to achieve signal-to-noise (S/N) ratio of 4 by 
using the {\it CIAO} tool {\it csmooth}. 
Sgr A* is marked near the center of the image. The bright X-ray feature
around Sgr A* is emission from a massive star cluster within the inner 
parsec of the Galactic center. The bright diffuse X-ray emission to the 
immediate east of Sgr A* is the SNR Sgr A East. As reported with previous 
{\it Chandra} observations by M02, X-ray emission from Sgr A East is centrally 
enhanced with no apparent shell-like features (Figure~\ref{fig:fig1}). The 
enhancements of the blue emission around the center of the SNR are remarkable. 
As we discuss in the next section, this enhanced hard X-ray emission is primarily 
from the Fe He$\alpha$ line emission. The outskirts of the SNR show soft X-ray 
emission. Also evident is red, soft X-ray emission extending toward the northern 
side of the SNR, the so-called ``plume'' (M02; Baganoff et al. 2003). M02 
suspected the existence of a high-velocity neutron star at the ``tip'' of the plume, 
which is physically associated with Sgr A East.  With the deep exposure, a hard 
pointlike source CXOGC J174545.5$-$285829 was indeed detected there 
\citep{muno03a}. This hard pointlike source might thus be the {\it predicted} 
high-velocity neutron star candidate, which might have also produced the bow 
shock-like plume: hereafter, we name this NS candidate the ``cannonball''. 
We discuss the {\it cannonball} in detail in \S~\ref{sec:ns} and 
\S~\ref{subsec:nsdisc}.

\subsection{\label{subsec:ewi} Equivalent Width Images}

In order to examine the overall distributions of the diffuse emission line 
features from Sgr A East, we construct {\it equivalent width} (EW) images for 
the detected atomic emission lines, following the method described in Park et al. 
(2004a). After removing all detected point sources from the broadband image 
(see Muno et al. [2003a] for the details of the point source detection), subband 
images for the line and continuum bandpasses were extracted for each spectral 
line of interest. These subband images were adaptively smoothed to achieve an S/N 
ratio of 3. The underlying continuum was calculated by logarithmically 
interpolating between images made from the higher and lower energy ``shoulder'' 
of each broad line. The estimated continuum flux was integrated over the selected 
line width and subtracted from the line emission. The continuum-subtracted line 
intensity was then divided by the estimated continuum on a pixel-by-pixel basis 
to generate the EW images for each element. In order to avoid noise in the EW maps 
caused by poor photon statistics near the CCD chip boundaries, we have set the EW 
values to zero where the estimated continuum flux is low. 
As discussed in Park et al. (2004a), contaminations from the cosmic X-ray 
background and the particle background, both of which have not been subtracted 
in the EW generation, are insignificant in the EW images. 
Although the adaptive smoothing may introduce spurious faint features, the bright, 
arcminute-scale features are confirmed by our spectral fits (see below) and are 
certainly real. Therefore, we focus on these bright features in order to 
qualitatively investigate the overall variation in the X-ray line emission of 
Sgr A East.

We present the Fe He$\alpha$ line ($E$ $\sim$ 6.6 keV) EW image of Sgr A East 
in Figure~\ref{fig:fig2}. The energy band selections for the line and continuum 
to generate this EW image are presented in Table~\ref{tbl:tab2}. The Fe He$\alpha$ 
line EW is strongly enhanced within an $\sim$40$^{\prime\prime}$ diameter region 
near the center of the SNR. This Fe EW distribution is consistent with the central 
concentration of the Fe abundance in Sgr A East as reported with the previous
{\it Chandra} and {\it XMM-Newton} observations (M02; S04). The high Fe EW at 
the center of the SNR is thus most likely caused by the enhanced Fe abundance 
there, displaying the distribution of Fe-rich stellar ejecta material. 
(Note: The bright X-ray emission at the position of Sgr A* is featureless in 
the Fe EW image. This is reasonable for the continuum-dominated X-ray spectrum 
from Sgr A* [Baganoff et al. 2003], which further supports the utility of
our EW images.) In contrast to the central Fe-rich region, the 
northern plume region of the SNR does not show an enhancement in Fe, which 
suggests that it is produced by emission from shocked ISM having metal abundances 
close to the average value at the Galactic center.

Unlike Fe, the EW maps of other elemental species S, Ar, and Ca do not show any 
significant spatial structure, and we do not explicitly present those EW images. 
The {\it uniform} EW distributions of other species are also consistent with the 
previously reported elemental abundance distributions across Sgr A East.

\subsection{\label{subsec:snrspec} X-Ray Spectra}

We perform a spatially-resolved spectral analysis of Sgr A East utilizing high 
angular resolution ACIS data. For the spectral analysis of our CTI-corrected data, 
we use the response matrices appropriate for the spectral redistribution of the 
CCD, as generated by Townsley et al. (2002b). Because of various roll-angles of 
the individual ACIS observations and the moderate angular extension of the SNR 
($\sim$3$'$), Sgr A East has been detected on different ACIS-I chips depending 
on the roll-angles. Observations taken with deep exposures between 2002 February 
and 2002 June (ObsID 2943 $-$ 3665 in Table~\ref{tbl:tab1}), however, had similar 
roll-angles, which represents $\sim$90\% of the total exposure. Small sections of 
the SNR have thus been detected on a single CCD for the majority of the 
observations.  For instance, although the bright, central Fe-rich region (i.e., 
``center'' region in Figure~\ref{fig:fig1}) was detected on either the ACIS-I1 or 
the ACIS-I3 depending on the roll-angles, $\sim$91\% of the total counts in the 
1 $-$ 8 keV band were detected on the ACIS-I1. We thus use detector responses 
appropriate for the chip positions of the source region on the ACIS-I1 for 
the ``center'' region spectrum. We used the same method for other regional spectra. 
We binned the data to contain a minimum of 30 $-$ 50 counts per channel for the 
spectral fittings. Even though we corrected the data for the CTI, there remain 
some residual effects of small gain shifts, which are usually insignificant in 
the spectral analysis. The gain shift at $\sim$1\% level may, however, confuse 
the spectral analysis of strong emission lines in a relatively narrow energy 
band such as Sgr A East spectrum containing strong Fe K lines between $E$ $\sim$ 
6.4 keV $-$ 6.9 keV. We inspected any residual gain shifts with the ACIS-I 
calibration data (ObsID 61184) taken close to our deepest observations (2002 May 25). 
We found that there were $\sim$0.1\% $-$ 1\% gain shifts depending on the detector 
positions. We thus included a ``gain fit'' in the initial spectral fittings and 
then adjusted the gain according to the best-fit gain shift. The adjusted gain 
shifts are typically small ($\sim$0.6\%), and are consistent with those from
the calibration data.    

The background estimates for the spectral analysis of Sgr A East are difficult
because X-ray background emission in the vicinity of Sgr A East shows complex 
spectral and spatial structure. We were particularly cautious about the 
background emission features from the neutral Fe K ($E$ $\sim$ 6.4 keV) and the
H-like Fe K ($E$ $\sim$ 6.96 keV) lines, because line emission from these species
is prevalent across the Galactic center region \citep{park04a,muno04a}. 
We considered several background regions which are source-free and represent the 
Fe line features. We eventually used the combined background emission from three 
regions which, we believe, adequately represent the ``average'' background spectrum 
for Sgr A East. The adopted background regions are presented in Figure~\ref{fig:fig3}. 
The spectral analysis and discussion in the following sections are based on this 
background selection. 
   
\subsubsection{\label{subsubsec:centerspec} Central Region}

We first extracted and fitted the spectrum of the $\sim$14$^{\prime\prime}$ 
$\times$ 24$^{\prime\prime}$ region containing the bright, Fe-rich emission 
at the core of the SNR (``center'' in Figure~\ref{fig:fig1} \& 
Figure~\ref{fig:fig2}). This region selection was also intended to include the 
central region with the highest broadband intensity, while excluding the 
chip-gap. Even with the small angular size of $\sim$20$^{\prime\prime}$, the 
``center'' spectrum contains significant photon statistics of $\sim$7000 counts, 
and clearly shows the strong Fe He$\alpha$ line at $E$ $\sim$ 6.6 keV 
(Figure~\ref{fig:fig4}). Since M02 and S04 consistently found that thermal plasma 
in a collisional ionization equilibrium (CIE) can adequately describe the Sgr 
A East spectrum, we fitted the ``center'' spectrum with a CIE model ({\it vmekal} 
in XSPEC) absorbed by interstellar gas (we have also fitted the spectrum with a 
non-equilibrium ionization [NEI] model and the best-fit ionization timescale 
was indeed high [$n_et$ $\sim$ 10$^{12}$ cm$^{-3}$ s], indicating a CIE plasma 
condition). The fitted absorption column was large ($N_H$ $\sim$ 15 $\times$ 
10$^{22}$ cm$^{-2}$), which was consistent with the Galactic center location of 
the SNR. The electron temperature ($kT$ $\sim$ 2 keV) was consistent with that 
reported by M02. The high Fe abundance of a few times solar was also in agreement 
with the results of M02 and S04. 

Although a single temperature CIE model may describe the observed spectrum, we 
recognized a residual ``hump'' at $E$ $\sim$ 7 keV above the best-fit model,
which appears to be a weak Fe Ly$\alpha$ line feature ($E$ $\sim$ 6.96 keV). 
In fact, the relatively poor fit ($\chi^2_{\nu}$ $\sim$ 1.4) was primarily 
caused by this excess emission. Our attempts to remove these residuals by 
adopting background subtraction from various regions (including the same 
background regions used by M02 and S04) were not successful. We thus conclude 
that this residual emission originates from the SNR. The detection of the 
Fe Ly$\alpha$ line is indeed statistically significant ($\sim$7$\sigma$).
This excess emission at $E$ $\sim$ 7 keV due to the H-like Fe line is 
important because an additional hot plasma component is required in order to 
fit this feature. The two-temperature CIE model can then improve the overall 
fit significantly ($\chi^2_{\nu}$ $\sim$ 1.0; Figure~\ref{fig:fig4}). In this 
two-component fit, we varied S, Ca, Ar, and Fe abundances while keeping them
in common for both components. All other species were fixed at solar abundance
because the contribution from those elemental species in the fitted energy band 
is negligible. We also assumed the same foreground column for both components. 
The results of this two-component spectral fit are presented in 
Table~\ref{tbl:tab3} and Table~\ref{tbl:tab4}. The fitted Fe abundance is high 
($\sim$5.8 solar) while those for S ($\sim$0.7), Ar ($\sim$1.8), and Ca 
($\sim$1.4) are relatively low.

The electron temperatures from the best-fit model are $kT$ $\sim$ 1 keV and 5 keV, 
which are consistent with the results from S04. Although the electron temperature 
for the hard component, $kT$ $\sim$ 5 keV, appears to be unusually high for an SNR 
spectrum, such a temperature is required to properly describe the detected 
Fe Ly$\alpha$ line feature. This highly ionized H-like Fe line emission was not 
detected by the early {\it Chandra} observations with short exposures (M02). 
It was, however, suggested by the {\it XMM-Newton} data (S04), and is now confirmed 
by our deep {\it Chandra} observations. S04, with the {\it XMM-Newton} data, also 
reported detection of ``neutral'' Fe line emission ($E$ $\sim$ 6.4 keV) from 
Sgr A East. Although more prominent in the peripheral regions of the SNR, the 
neutral Fe K line was reportedly detected in the central regions of the SNR with 
{\it XMM-Newton} (photon flux $\sim$ 3 $\times$ 10$^{-6}$ photons cm$^{-2}$ s$^{-1}$ 
within 28$^{\prime\prime}$ radius; S04). Our deep, high angular resolution 
{\it Chandra} data do not require such a neutral Fe line feature in order to 
describe the observed spectrum. We place a 2$\sigma$ upper limit of 1.1 $\times$ 
10$^{-6}$ photons cm$^{-2}$ s$^{-1}$ on the 6.4 keV Fe line photon flux.  

\subsubsection{\label{subsubsec:plumespec} ``Plume'' Regions}

\begin{figure}[]
\figurenum{2}
\centerline{{\includegraphics[width=0.8\textwidth]{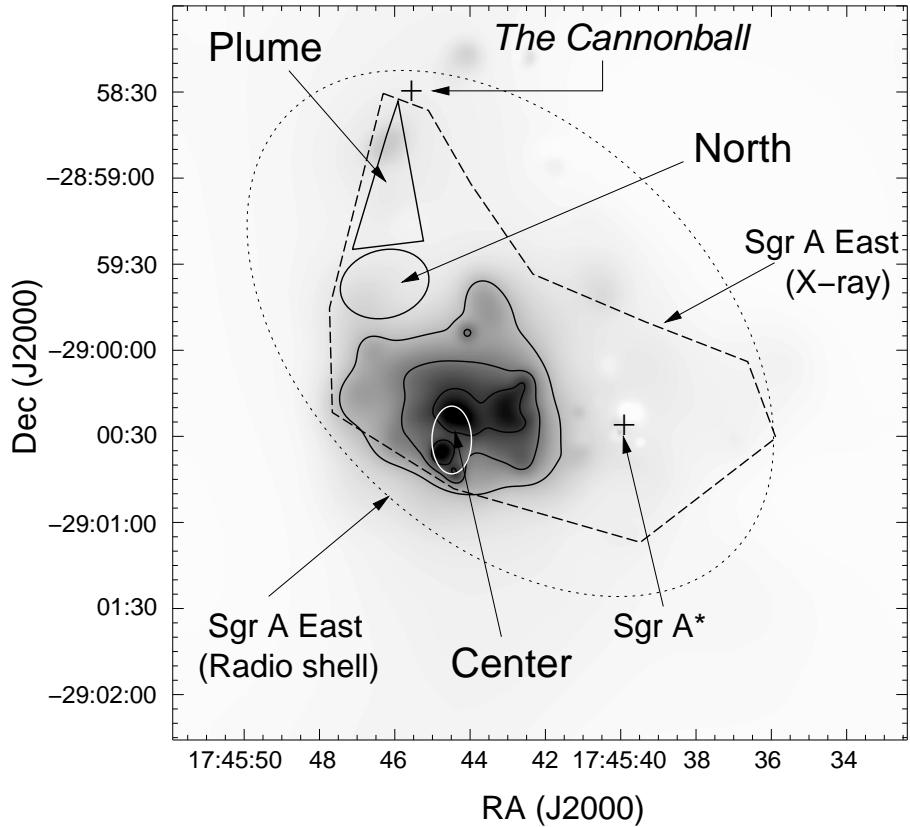}}}
\figcaption[]{Gray-scale EW image of Fe He$\alpha$ line ($E$ $\sim$ 6.6 keV) 
from Sgr A East. Linear gray-scales (with darker gray-scales correspondign to 
higher intensities) range from 200 eV to 2500 eV. Three contours represent 800 eV, 
1400 eV, and 2000 eV.  
The line and continuum images were adaptively smoothed prior to calculation 
of the EW. 
Some key features within the FOV and regions used for the spectral analysis are marked. 
\label{fig:fig2}}
\end{figure}

\begin{figure}[]
\figurenum{3}
\centerline{\includegraphics[angle=0,width=0.5\textwidth]{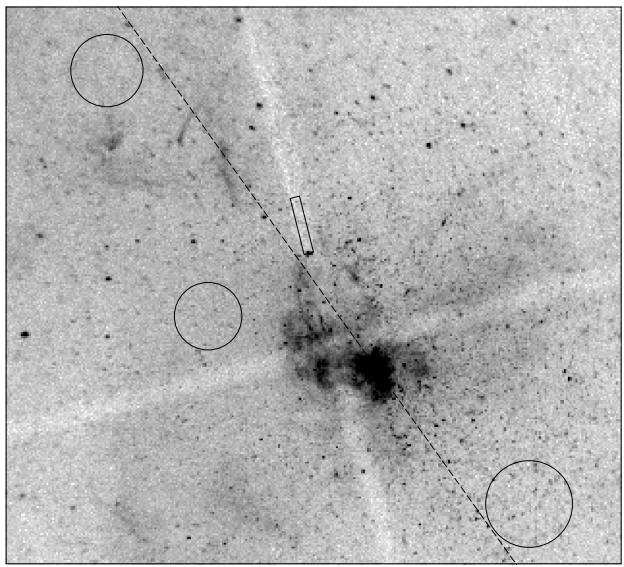}}
\figcaption[]{The {\it Chandra}/ACIS image of Sgr A East and its vicinity.
Three circles are regions where the background emission was extracted for the
SNR spectral analysis. The small rectangular box along the ACIS-I chip-gap is 
the background region for the X-ray spectrum of the {\it cannonball}. The 
dashed diagonal line shows the Galactic plane through Sgr A*. Darker gray scales
are higher intensities. This broadband (1.5 $-$ 8 keV) image has been rebinned
by 4 $\times$ 4 pixels and was not exposure-corrected. Detected point sources
have not been removed in this image, although the actual background spectra
were extracted after removing all detected point sources.
\label{fig:fig3}}
\end{figure}

\begin{figure}[]
\figurenum{4}
\centerline{\includegraphics[angle=-90,width=0.5\textwidth]{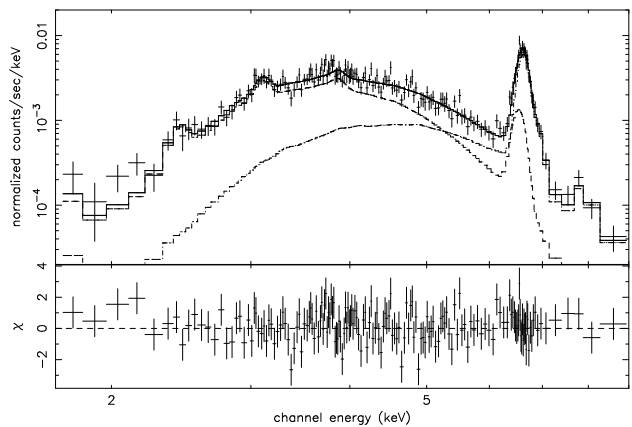}}
\figcaption[]{The {\it Chandra}/ACIS spectrum of the Fe-rich core of
Sgr A East (``center'' region). The best-fit two-temperature thermal
plasma model is overlaid.
\label{fig:fig4}}
\end{figure}

\begin{figure}[]
\figurenum{5}
\centerline{\includegraphics[angle=-90,width=0.5\textwidth]{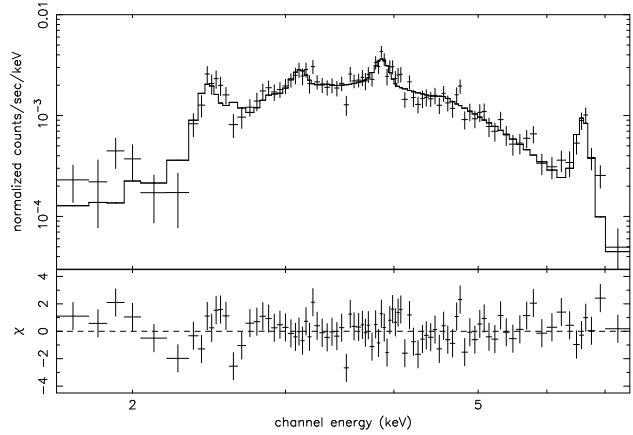}}
\figcaption[]{The {\it Chandra}/ACIS spectrum of the ``plume'' region.
The best-fit single temperature thermal plasma model is overlaid.
\label{fig:fig5}}
\end{figure}

We extract the spectrum of the plumelike feature from a region between the Fe-rich 
center and the {\it cannonball} (``plume'' region in Figure~\ref{fig:fig1} \& 
Figure~\ref{fig:fig2}). The spectrum contains $\sim$5300 counts and can be fitted 
with a single-temperature ($kT$ $\sim$ 1.3 keV) CIE plasma model 
(Figure~\ref{fig:fig5}). In this fit, the elemental abundances for S, Ar, Ca, and 
Fe were varied freely, while abundances for other species were fixed at solar
(Table~\ref{tbl:tab3} and Table~\ref{tbl:tab4}). The Fe He$\alpha$ line is 
significantly weaker (e.g., EW $\sim$ 800 eV) than the ``center'' region spectrum 
(e.g., EW $\sim$ 2500 eV). The Fe abundance is also considerably lower than the 
central region, and is consistent with solar. These results indicate that the 
plume is emission primarily from a shocked ISM rather than from metal-rich ejecta. 
In contrast to the ``center'' region, hard excess emission at $E$ $\sim$ 7 keV 
due to the Fe Ly$\alpha$ line feature is not substantial in the plume region: 
i.e., the line detection is only marginal ($\sim$2.5$\sigma$), and the line flux 
is an order of magnitude lower than that from the ``center'' region. In fact, 
the statistical improvement with a two-temperature plasma model is insignificant 
(e.g., $\chi^2_{\nu}$ $\sim$ 1.1 vs. 1.2). The fitted metal abundances also 
unchange when a two-temperature model is used. Although there appears to be a 
small contribution from the high temperature ($kT$ $\sim$ 5 keV) component, the 
overall spectrum from the plume region is thus dominated by the soft component 
(kT $\sim$ 1 keV). Therefore, we hereafter discuss the plume spectrum based on 
the single-temperature model, as presented in Table~\ref{tbl:tab3} and 
Table~\ref{tbl:tab4}. 

\begin{figure}[]
\figurenum{6}
\centerline{\includegraphics[angle=-90,width=0.5\textwidth]{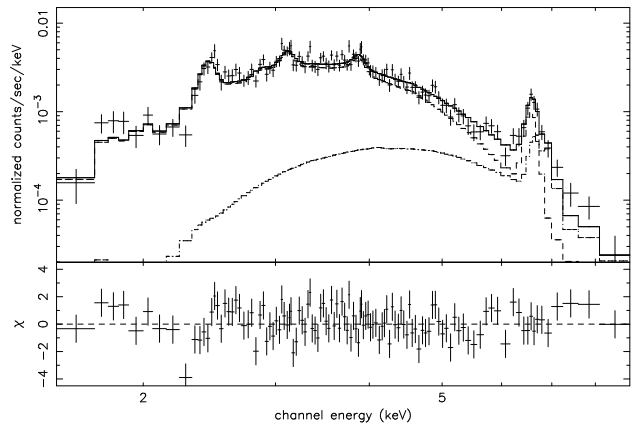}}
\figcaption[]{The {\it Chandra}/ACIS spectrum of of the ``north'' region.
The best-fit two-temperature thermal plasma model is overlaid.
\label{fig:fig6}}
\end{figure}

We then extract the spectrum from a region between ``center'' and ``plume'' 
(``north'' region in Figure~\ref{fig:fig1} \& Figure~\ref{fig:fig2}). This 
regional spectrum contains $\sim$7600 counts. The Fe He$\alpha$ line is 
substantially weaker than the ``center'' region, similar to the 
``plume'' spectrum. On the other hand, as in the ``center'' region spectrum, 
hard excess emission at $E$ $\ga$ 7 keV was noticeable for the ``north'' region 
spectrum when fitted with a single-temperature model. We thus fit this regional 
spectrum with a two-temperature CIE plasma model in the same way as 
for the ``center'' region (Figure~\ref{fig:fig6}). The fitted parameters from 
the two-temperature model are presented in Table~\ref{tbl:tab3} and 
Table~\ref{tbl:tab4}. The fitted parameters are generally consistent with those 
from the central region.
It is, however, notable that the Fe abundance in the ``north'' region is 
significantly lower than that of the ``center'' region. As with the other regions,
we found that the 6.4 keV line is not evident in the observed ``north'' region 
spectrum, with an upper limit to the line flux of 3.7 $\times$ 10$^{-7}$ photons 
cm$^{-2}$ s$^{-1}$ (2$\sigma$).

\section{\label{sec:ns} CXOGC J174545.5$-$285829: {\it The Cannonball}}

\subsection{\label{subsec:nsimage} X-Ray Morphology}

\begin{figure}[]
\figurenum{7}
\centerline{\includegraphics[angle=0,width=0.8\textwidth]{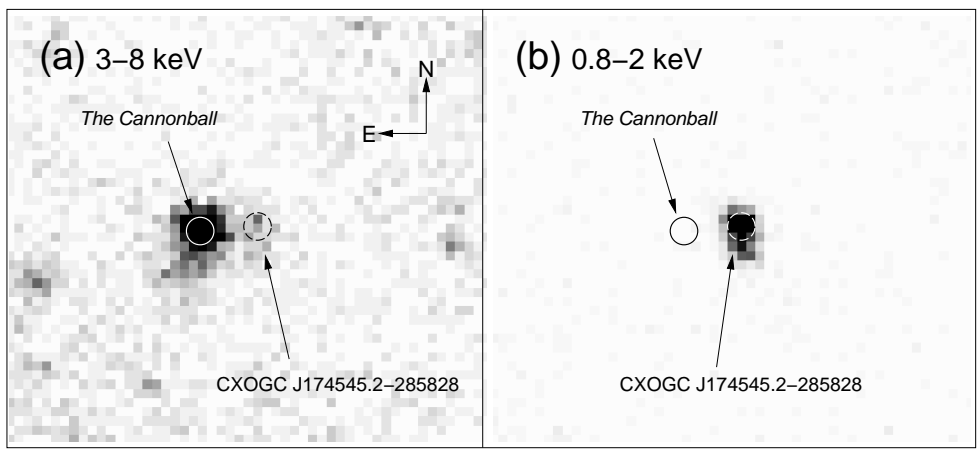}}
\figcaption[]{The {\it Chandra}/ACIS image of the {\it cannonball}
and CXOGC J174545.2$-$285828. (a) The hard band (3 $-$ 8 keV) image.
(b) The soft band (0.8 $-$ 2 keV) image. Darker gray-scales are higher intensities.
\label{fig:fig7}}
\end{figure}

M02 speculated that a high-velocity neutron star associated with the SNR Sgr A 
East might exist at the ``tip'' of the plume.  With deep observations, a hard 
pointlike source was indeed detected at the edge of the plume and designated as 
CXOGC J174545.5$-$285829 (the so-called {\it cannonball}) by Muno et al. (2003a). 
Figure~\ref{fig:fig7} presents the soft and hard band images of the {\it cannonball}. 
In fact, at the edge of the plume, there are two pointlike sources separated by 
$\sim$3$^{\prime\prime}$: the {\it cannonball} detected exclusively above 
$E$ $\sim$ 3 keV, and a soft source CXOGC J174545.2$-$285828 detected mostly 
below $E$ $\sim$ 2 keV (Figure~\ref{fig:fig7}). The soft source CXOGC 
J174545.2$-$285828 exhibits long-term variability \citep{muno04b}, and has an 
optical counterpart in the USNO-B1.0 catalog (\#0610-0600649, m$_B$ = 15.27), 
both of which suggest that this soft source is a foreground star.

\begin{figure}[]
\figurenum{8}
\centerline{\includegraphics[angle=0,width=0.9\textwidth]{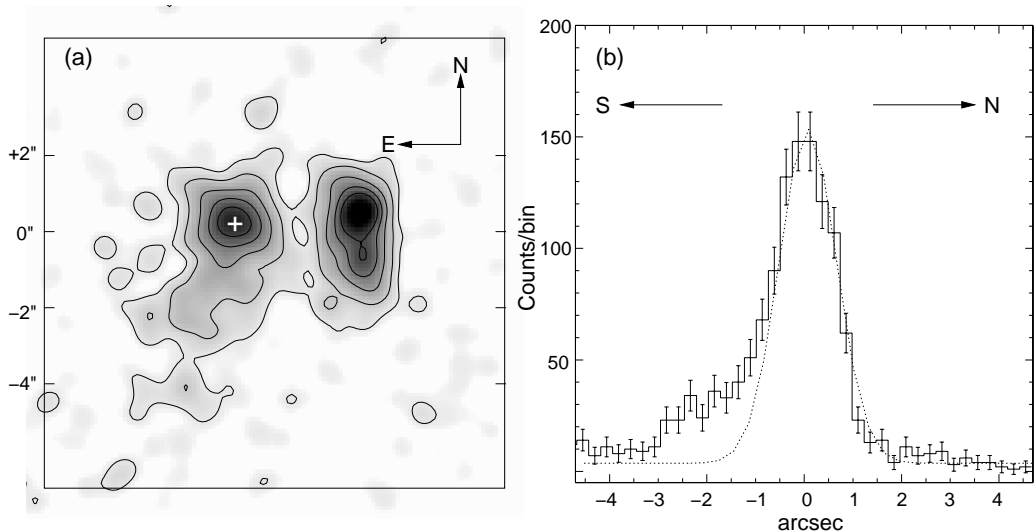}}
\figcaption[]{
(a) The deconvolved {\it Chandra}/ACIS images of the {\it cannonball} (left
source) and CXOGC J174545.2$-$285828 (right source). 
The image has been smoothed with a Gaussian of $\sigma$ = 0$\farcs$25 for the 
purpose of display.
(b) The source intensity profiles projected in the north-south direction taken
from the raw ACIS image. The histogram represents the {\it cannonball}, and
the projected PSF at the source position is overlaid with a dashed curve.
\label{fig:fig8}}
\end{figure}

Figure~\ref{fig:fig7} reveals that the {\it cannonball} may not be truly 
pointlike, but is apparently extended: i.e., a cometary shape with a bright, 
pointlike ``head'' and a ``tail'' extending generally southward, which is 
roughly tracking back to Sgr A East. Although this source was detected 
$\sim$2$\farcm$3 off-axis, the apparent extended morphology is not caused by the 
point spread function (PSF). In Figure~\ref{fig:fig8}b, the one-dimensional
source intensity distribution in the north-south direction is compared to that
of the off-axis PSF at the same position. The extended nature of the 
{\it cannonball} toward the 
south is unambiguously revealed. In Figure~\ref{fig:fig8}a, we present a 
PSF-deconvolved source image, using a maximum-likelihood algorithm 
\citep{rich72,lucy74}. For comparison, the deconvolved image of the soft source 
CXOGC J174545.2$-$285828 (the source in the right-hand side of the image) is 
also presented. The {\it cannonball} shows a significantly extended 
structure which confirms the cometary morphology suggested by the raw image and 
the projected intensity profile. The soft source, in contrast, appears to be
resolved further into two pointlike sources. The photon statistics of the two 
sources, {\it the cannonball} and CXOGC J174545.2$-$285828, are similar 
($\sim$1000 $-$ 1200 counts), and thus the morphological difference between the 
two is unlikely a statistical artifact. The angular extent of the {\it 
cannonball}'s tail is $\sim$3$^{\prime\prime}$, which corresponds to a projected 
physical size of $\sim$0.1 pc (Hereafter, we use a distance of 8 kpc to the 
Galactic center [Reid 1993]). 

\subsection{\label{subsec:nsspec} X-Ray Spectrum}

The {\it cannonball} has been detected on various detector positions of the 
ACIS-I array because of the different roll-angles for the individual observations. 
These various detector positions include on-chip and chip-gap positions as well 
as different ``rows'' within the CCD. We thus use detector responses appropriate 
for each observation to average over all the observations for the spectral 
analysis of the combined data. For six observations (ObsIDs 242, 1561b, 2951, 
2952, 2953, and 2954) in which the source was detected on an on-chip position, 
the source photons are relatively evenly spread over the ACIS-I0 (46\% of the 
total) and the ACIS-I3 (54\% of the total). We thus made a weighted sum (by 
detected counts) of the redistribution matrix function (RMF) files appropriate 
for the source locations of each CCD. The corresponding ancilliary response 
functions (ARF) were also averaged by weighting the counts detected in the 
individual observations. With the other five observations (ObsIDs 2943, 3663, 
3392, 3393, and 3665), nearly all of the photons (90\% of the total) were 
detected on the ACIS-I0. We thus use the RMF appropriate for the source detector 
positions on the I0 chip. The ARF files for these five observations were also 
averaged by weighting the counts detected in the individual observations.  

\begin{figure}[]
\figurenum{9}
\centerline{\includegraphics[angle=-90,width=0.5\textwidth]{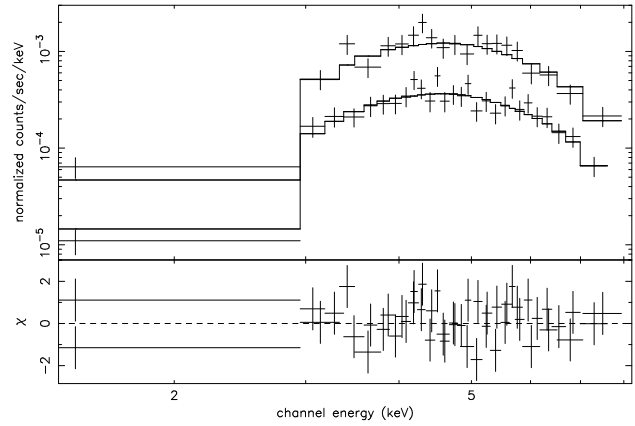}}
\figcaption[]{The {\it Chandra}/ACIS spectrum of the {\it cannonball}.
The upper plot is the on-chip spectrum and the lower plot is the chip-gap
spectrum. The best-fit power law model is overlaid.
\label{fig:fig9}}
\end{figure}

The source spectrum of the {\it cannonball} is extracted from an $\sim$1$\farcs$8 
radius circular region centered on the source position. We tested the background 
spectrum separately from three regions: i.e., an annulus of inner radius 
$\sim$5$^{\prime\prime}$ and outer radius $\sim$6$\farcs$5 centered on the 
source position, a circular region of radius $\sim$11$^{\prime\prime}$ in the 
northeast side of the source, and a 10$^{\prime\prime}$ $\times$ 60$^{\prime\prime}$ 
box region along the CCD chip-gap (for ObsIDs 2943, 3663, 3392, 3393, and 3665) to
the north of the source. The choice of these background regions resulted in no
significant differences in the spectral fits. In the following discussion of 
the spectral analysis, we assume the rectangular background region along the 
chip-gap (Figure~\ref{fig:fig1} \& Figure~\ref{fig:fig3}), because this 
background region consistently represents the source ``flux variation'' caused
by the on- and off-chip positions. We binned the source spectrum in order to 
contain 20 or more counts per channel. The spectrum can best be fitted with 
a power law (Figure~\ref{fig:fig9}) and the best-fit parameters are presented 
in Table~\ref{tbl:tab5}. Because of the relatively low photon statistics 
($\sim$1000 counts), the spectrum may also be fitted with thermal plasma models. 
The best-fit electron temperature is remarkably high ($kT$ $\sim$ 25 keV) 
with extremely low metal abundances. This high temperature might not be entirely
unreasonable for sources like cataclysmic variables. Although unlikely, a thermal
origin of the X-ray spectrum of the {\it cannonball} may thus not be
completely ruled out. 

\section{\label{sec:disc} DISCUSSION}

\subsection{\label{subsec:nsdisc} The {\it Cannonball}: A High-Velocity Neutron 
Star?}

The spectral and morphological characteristics of the {\it cannonball} are very 
unusual among the sources detected within the ACIS FOV of the Galactic center: 
i.e., it is a bright source with a hard, continuum-dominated X-ray spectrum
having no Fe line emission. This source also shows no long-term
variability \citep{muno04b}. The {\it cannonball} is one of only two sources, 
out of $\sim$2300 Galactic center X-ray sources cataloged by Muno et al. (2003a), 
which share such unusual X-ray characteristics. We discuss this unique source in 
the following sections. The observed X-ray nature of the {\it cannonball} 
suggests an identification as a high-velocity neutron star in the Galactic center. 
Based on its proximity to Sgr A East, we therefore propose that the 
{\it cannonball} may be a candidate neutron star born from the SN explosion 
which also produced the SNR, Sgr A East.

\subsubsection{\label{subsubsec:nsmorspec} Spectrum and Morphology}

The X-ray spectrum of the {\it cannonball} is best-fitted with a power law 
($\Gamma$ $\sim$ 1.6), which is typical for nonthermal synchrotron emission from
a neutron star's magnetosphere. The implied high foreground absorption is 
consistent with that of Sgr A East, in agreement with a Galactic center location 
nearby the SNR. Assuming a Galactic center location, the estimated X-ray luminosity 
(L$_X$ $\sim$ 3.1 $\times$ 10$^{33}$ ergs s$^{-1}$) is also typical for a pulsar 
and/or pulsar wind nebula (PWN) (e.g., Becker \& Aschenbach 2002).

The deep {\it Chandra} images unambiguously reveal a cometary morphology of the
{\it cannonball}. The tail is faint (only $\la$100 photons), and the estimated 
physical size is $\sim$0.1 pc.
This tail size is consistent with cometary tails detected from other Galactic 
high-velocity pulsars such as the Geminga pulsar \citep{cara03}, PSR B1853+01 
\citep{petre02}, and PSR J1509$-$5850 (Sanwal et al. in preparation). The tail
points roughly toward the south, consistent with the direction of the center
of Sgr A East, given its limited photon statistics. The sky position of the 
source is also interesting: i.e., the {\it cannonball} is located, in projection, 
just {\it interior} to the radio shell boundary of Sgr A East while sitting on 
the ``tip'' of the X-ray plume. These X-ray morphologies are suggestive of a 
high-velocity neutron star, moving toward the north, for the origin of the 
{\it cannonball}. Considering an SNR age of $\sim$5000 $-$ 10000 yr (Mezger et al. 
1989; Uchida et al. 1998; M02; S04) and the angular separation of $\sim$2$'$ 
between the source and the ``center'' of the SNR, a velocity of $v$ $\sim$ 
455 $-$ 912 km s$^{-1}$/sin$\beta$, where $\beta$ is the angle between the line 
of sight and the actual traveling direction of the neutron star, is implied. 
This velocity range is in good agreement with that of typical high-velocity 
pulsars in the Galaxy \citep{cordes98}. 

We may make an independent estimate of the velocity of the candidate neutron star
by assuming that the plume is a bow-shock produced as the high-velocity neutron star
encounters the ISM. Based on the conical shape of the plume as seen by the ``red'' 
emission feature in Figure~\ref{fig:fig1}a, we estimate an opening angle of 
$\theta$ $\sim$ 53$^{\circ}$. The Mach number Ma = [sin($\theta$/2)]$^{-1}$ is 
thus $\sim$2.2. The velocity of the candidate neutron star is then $v$ = $c_s$ 
Ma = $\sim$880 km s$^{-1}$. In this estimate, we assumed a sound speed of the 
ambient gas $c_s$ = ($\gamma$$kT$/$\mu$$m_p$)$^{1\over2}$ $\sim$ 400 km s$^{-1}$ 
where $\gamma$ = 5/3 for a monatomic, adiabatic gas, $\mu$ = 1 for protons, $m_p$ 
is the proton mass, and $kT$ $\sim$ 1 keV plasma in the plume region. We note that, 
in the velocity estimation from the Mach number, we used the projected opening angle 
for the bow-shock. If the inclination angle of the bow-shock was significant, 
the actual, de-projected opening angle could have been smaller, thus higher velocity 
would have been derived. Nonetheless, the velocity estimates from the two methods
are in plausible agreement.
   
A third velocity estimate can be derived by assuming that this NS candidate
produced a PWN which has reached pressure equilibrium with the ISM.
Assuming that the X-ray spectrum of the {\it cannonball} is primarily from
a PWN of the neutron star, we may derive the rotational spin-down energy loss 
$\dot{E}$ by using an empirical relationship between $\dot{E}$ and the power law 
photon index of the PWN, $$\Gamma_{PWN} = 2.36 - 0.021 \dot{E}_{40}^{-{1\over2}},$$
where  $\dot{E}_{40}$ is the spin-down power in units of 10$^{40}$ ergs s$^{-1}$
\citep{got03}. The best-fit photon index of $\Gamma_{PWN}$ = 1.6 implies $\dot{E}$ 
$\sim$ 7.6 $\times$ 10$^{36}$ ergs s$^{-1}$. We may assume a pressure balance 
between the ram pressure of the PWN, $P_{PWN}$ = $\dot{E}$/(4$\pi$cR$^2$) where R 
is the PWN radius for a spherical geometry, and the thermal pressure of the plume 
region, $P_{th}$ = 2$n_e$$kT$. The best-fit volume emission measure ($EM$) from 
the plume region implies an electron density of $n_e$ $\sim$ 7.4 $f^{-{1\over2}}$ 
cm$^{-3}$, where $f$ is the X-ray emitting volume filling factor (we assumed a 
half-conical volume with a circular base of radius $\sim$ 25$^{\prime\prime}$ and 
a height $\sim$ 50$^{\prime\prime}$ for the ``plume'' region). The best-fit 
thermal plasma model for the plume region then implies $P_{th}$ (= $P_{PWN}$) 
$\sim$ 3.1 $\times$ 10$^{-8}$ ergs cm$^{-3}$. (The PWN radius is accordingly
estimated to be R $\sim$ 2.5 $\times$ 10$^{16}$ cm, which corresponds to 
$\sim$0$\farcs$2 at $d$ $\sim$ 8 kpc. This small R is consistent with the 
pointlike detection of the ``head'' of the source by the ACIS.) $P_{PWN}$ may 
be equivalent to the external ram pressure $\epsilon$$_k$ = $\rho$$v^2$ ($\rho$ 
is the mass density of the ISM derived from the plume region spectrum). The 
velocity is then derived to be $v$ $\sim$ 550 km s$^{-1}$. We note that there 
were a number of assumptions in these estimates, e.g., the pressure balance 
$P_{PWN}$ = $P_{th}$ = $\epsilon$$_k$, the $\dot{E}$ estimate from the best-fit 
photon index of the PWN spectrum, and the bow-shock origin for the plume. 
Considering various assumptions and embedded uncertainties in the above
three approaches for the velocity estimates, the agreement among all three results 
are remarkable. The consistency among these independent estimates of the velocity 
thus lends all of them additional credibility.

\subsubsection{\label{subsubsec:nslc} Temporal Characteristics}

No evidence for long-term (observation by observation) variability of the
{\it cannonball} was detected \citep{muno04b}. The proximity of this source to 
the chip-gap during the long series of observations makes it difficult to
determine whether there is short-term (within individual observation periods)
variability. 
With a 2$-$10 keV band flux of $\sim$1.9 $\times$ 10$^{-13}$ ergs cm$^{-2}$ 
s$^{-1}$, the {\it cannonball} is one of the brightest sources detected in 
the ACIS FOV: i.e., the 5th brightest in terms of the photon flux, and
the 18th brightest in the number of counts among $\sim$2300 cataloged Galactic 
center sources \citep{muno03a}. According to Muno et al. (2004b), $\sim$80\% of 
the short-term variables have a {\it maximum} photon flux lower than the photon 
flux of the {\it cannonball}. Also, $\sim$94\% of the long-term variables 
have a {\it maximum} flux lower than the {\it cannonball} \citep{muno04b}. 
For comparison, the nearby soft source CXOGC J174545.2$-$285828 
with similar photon statistics was identified as a long-term variable 
\citep{muno04b}. The non-variability of the {\it cannonball} is thus unlikely 
a statistical artifact, but should correspond to the true nature of this object. 
The constant lightcurve with a long time-basis of $\sim$3 yr, particularly at 
the ``high'' flux level of this source, is remarkably unusual among the Galactic 
center sources. This non-variability over a few years indicates that the 
{\it cannonball} is unlikely a background AGN, and supports a neutron star 
identification. In fact, based on the logN-logS relation for the contribution 
from extragalactic sources in the Galactic center \citep{muno03a}, a low 
probability of $\sim$1.2 $\times$ 10$^{-4}$ for a detection of a background 
extragalactic source at the flux level of the {\it cannonball} within the 
``plume'' region is implied. 
 
The discovery of a pulsar at the position of the {\it cannonball} would 
conclusively identify this source as a neutron star. Our deep {\it Chandra} 
observations, however, used the standard 3.2 s time frame which may not be adequate 
for a pulsar search. We, therefore, searched for pulsations in a 40-ks  
archival {\it XMM-Newton}/EPIC observation pointed at Sgr A* (observation
sequence number 0111350101). The EPIC/PN instrument has a relatively short frame 
time of 73.4 ms even in full-window mode, so that our pulsar search was sensitive 
down to periods of $\sim$147 ms. Because of the poor angular resolution of 
the EPIC instrument, the two sources, the {\it cannonball} and CXOGC 
J174545.2$-$285828, are not resolved. In order to discriminate contamination from 
the nearby soft source CXOGC J174545.2$-$285828, we extract the source photons 
only from the hard band (3 $-$ 10 keV). We then detect $\sim$460 counts from 
the {\it cannonball}. With these {\it XMM-Newton} 
data, we do not detect significant evidence of a pulsar. We place an upper limit 
of $\sim$40\% on the pulsed-fraction between 5 $\times$ 10$^{-5}$ Hz and 6.8 Hz. 
If the {\it cannonball} is a pulsar associated with Sgr A East and thus has 
an age of $\sim$10000 yr, it may have a Vela pulsar-like periodicity of 
$\sim$100 ms. This pulsation period is beyond the detectability of the used
{\it XMM-Newton} time resolution. The presence of a pulsar for this neutron star 
candidate cannot thus be ruled out yet. Follow-up X-ray and/or radio observations 
with high time resolution instruments would therefore be helpful for the pulsar 
search for this NS candidate. 

\subsection{\label{subsec:snrdisc} Sgr A East: Fe Ejecta Mass}

The enhancement of strong Fe K line emission at the center of the SNR is 
remarkable. This Fe line emission is most likely from the Fe-rich stellar ejecta 
from the progenitor, which was heated by the reverse shock. Assuming that the 
Fe-rich central region represents the bulk of the total Fe mass ejected from
the progenitor star, this Fe-rich material provides a useful opportunity to 
investigate the progenitor mass, and thus the SN explosion type of Sgr A East. 

In order to estimate the total Fe ejecta mass, we first consider a simple 
geometry in which {\it all} Fe ejecta material is concentrated within the central 
$\sim$40$^{\prime\prime}$ diameter region where the Fe EW is the highest
(e.g., Fe EW $\ga$ 1400 eV in Figure~\ref{fig:fig2}). We estimate the electron
density based on the measured $EM$ from the spectral fit of the ``center'' region. 
We use a cylindrical geometry with an elliptical cross-section (major and minor 
axis radii of 12$^{\prime\prime}$ and 7$^{\prime\prime}$, respectively) for 
the X-ray emitting volume of the ``center'' region. A path-length of 1.55 pc 
was assumed, corresponding to $\sim$40$^{\prime\prime}$ angular extent of the
Fe enhancement at the center of the SNR. The X-ray emitting volume of $V$ = 
1.82 $f$ $\times$ 10$^{55}$ cm$^3$ is thus used for the density estimate. 
We, for simplicity, assume a ``pure'' Fe-ejecta case for the estimate of the 
electron density: i.e., {\it all} electrons originate from the ionized Fe. 
Considering the dominant ionization states presented by two-temperature plasma, 
we assume electron-Fe ion density ratio of $n_e$ = 24$n_{Fe}$ for the He-like Fe. 
We also use the best-fit Fe abundance of 5.8 $\times$ solar for each of the soft 
and the hard components. Considering the large difference in the plasma 
temperatures of the two components, they are unlikely ``co-spatial'' (e.g., S04). 
We thus assume that each component occupies a separate volume and maintains a 
pressure equilibrium: 
i.e., $n_{es}$T$_{s}$ = $n_{eh}$T$_h$ and $f$ = $f_{s}$ + $f_h$, where $n_{es}$ 
and $n_{eh}$ are electron densities, T$_{s}$ and T$_h$ are electron temperatures, 
and $f_{s}$ and $f_{h}$ are volume filling factors for the soft and the hard 
component, respectively. The best-fit electron temperatures ($kT$ = 1.05 keV and 
5.28 keV) correspond to $f_{s}$ = 0.48$f$ and $f_{h}$ = 0.52$f$. The best-fit 
$EM$s then imply Fe ion densities of $n_{Fes}$ $\sim$ 0.096$f^{-{1\over2}}$ 
cm$^{-3}$ and $n_{Feh}$ $\sim$ 0.019$f^{-{1\over2}}$ cm$^{-3}$, where $n_{Fes}$ and 
$n_{Feh}$ are Fe ion densities for the soft and the hard component, respectively.
If we assume a spherical volume with a diameter of $\sim$1.55 pc and with a 
dominant isotope $^{56}$Fe, the {\it total} Fe mass, M$_{Fe}$ = 56($n_{Fes}$ + 
$n_{Feh}$)$m_pV$, is then estimated to be M$_{Fe}$ $\sim$ 0.15$f^{1\over2}$ 
M$_{\odot}$. Because of our ``pure'' Fe ejecta assumption, this Fe mass is an 
upper limit. 

The standard SN nucleosynthesis models indicate that M$_{Fe}$ is $\sim$0.5M$_{\odot}$ 
$-$ 0.8M$_{\odot}$ for the Type Ia SNe (e.g., Nomoto et al. 1997a). The Type II 
models show a wide range of Fe mass depending on the progenitor masses: e.g., 
M$_{Fe}$ $\sim$ 0.15M$_{\odot}$ for a 13M$_{\odot}$ $-$ 15M$_{\odot}$ progenitor, 
and M$_{Fe}$ $\sim$ 0.05M$_{\odot}$ $-$ 0.08M$_{\odot}$ for a 18M$_{\odot}$ $-$ 
70M$_{\odot}$ progenitor (e.g., Nomoto et al. 1997b). The estimated upper limit on 
the Fe mass for Sgr A East (M$_{Fe}$ $\sim$ 0.15M$_{\odot}$) is thus consistent 
with a Type II origin from a 13M$_{\odot}$ $-$ 15M$_{\odot}$ progenitor. If our 
Fe mass estimates were significantly affected by the ISM contribution, the 
progenitor mass could be larger. 

Although the total Fe mass, as estimated assuming a simple spherical geometry,
provided a useful constraint on the progenitor mass, we may also entertain
a contribution to the total Fe mass of the SNR from the region where Fe is marginally 
enhanced. Inclusion of this additional Fe mass may provide an even more conservative 
limit on the progenitor mass. The Fe EWs are moderately enhanced in an extended 
region just outside of the $\sim$40$^{\prime\prime}$ Fe core (i.e., the region 
between the Fe EW of 800 eV and 1400 eV contours in Figure~\ref{fig:fig2}). 
This region may be represented by a ``spherical shell'' with an outer diameter 
of $\sim$1$'$ and an inner diameter of $\sim$40$^{\prime\prime}$. The Fe abundance 
in this region appears to be $\sim$3. We thus consider the contribution from this 
moderately Fe-rich region to the total Fe mass of the SNR. We assume that the 
path-length through the outer shell region (foreground and background of the 
Fe-core in projection) is $\sim$50\% of that for the inner Fe core for the adopted 
geometry. With the pure ejecta assumption, we consider that $n_e$ and $n_{Fe}$ 
are simply proportional to the Fe abundance ratio between the two regions (i.e., 
5.8 for the inner core and 3 for the outer shell). The contribution from the shell 
region to the measured $EM$ is then $\sim$10\%. We thus use the fitted $EM$s 
from the ``center'' region by assuming an $\sim$90\% contribution from the inner 
Fe core ($\sim$1.55 pc diameter, corresponding to $\sim$40$^{\prime\prime}$ angular 
size at 8 kpc) and an $\sim$10\% contribution from the outer shell ($\sim$0.39 pc 
thickness, corresponding $\sim$10$^{\prime\prime}$ angular size at 8 kpc). 
We also assume the same fractional volume filling factors ($f_{s}$ = 0.48$f$ and 
$f_{h}$ = 0.52$f$) as above. The derived total Fe densities are $n_{Fe}$ $\sim$ 
0.107$f^{-{1\over2}}$ cm$^{-3}$ for the inner Fe core and $n_{Fe}$ $\sim$ 
0.038$f^{-{1\over2}}$ cm$^{-3}$ for the outer shell. We then estimate the total Fe 
mass from the $\sim$1$'$ diameter region, M$_{Fe}$ $\sim$ 0.27$f^{1\over2}$ 
M$_{\odot}$, which is $\sim$80\% larger than that derived merely from an 
$\sim$40$^{\prime\prime}$ diameter spherical core. This Fe mass is again a 
conservative upper limit because of the pure Fe ejecta assumption.  
We find that this upper limit of the total Fe mass of Sgr A East is still 
considerably lower than that expected from Type Ia models.
Based on these derived Fe mass limits, we suggest that Sgr A East was produced 
likely by a core-collapse SN explosion of a massive progenitor star. A core-collapse 
origin for the SNR Sgr A East then supports the proposed SNR-NS association between 
Sgr A East and the {\it cannonball}. 

We note, however, that a Type Ia origin for Sgr A East might not be completely 
ruled out, yet. Considering the relatively old age of the SNR, one might speculate 
that a significant fraction of the Fe-rich ejecta has already been intermixed with 
the ambient ISM. The apparent lack of abundance enhancements from other high-Z
elemental species may support this scenario. In such a case, the derived upper
limits of Fe ejecta mass could be underestimates. If the amount of ``missing'' 
Fe ejecta mass was {\it sufficiently} large ($\sim$several times larger than
the observed Fe ejecta mass), the ``true'' total Fe ejecta mass might have been
large enough to be consistent with a Type Ia origin for Sgr A East.

\subsection{\label{subsec:line} Comments on Fe Lines}

The high angular resolution of the ACIS coupled with the deep exposure should be 
useful to investigate some controversial results between previous {\it Chandra} 
(M02) and {\it XMM-Newton} (S04) observations of Sgr A East: i.e., the {\it
XMM-Newton} data reported line emission from highly-ionized H-like Fe and 
low-ionization state ``neutral'' Fe as well as the strong Fe He$\alpha$ line (S04),
while the previous {\it Chandra} data showed only strong Fe He$\alpha$ line
without H-like and neutral Fe lines (M02). These differences in the detected
Fe line features resulted in discrepancies in the plasma temperatures ($kT$ $\sim$
2 keV for the {\it Chandra} data vs. $kT$ $\sim$ 1 keV and 4 keV for the {\it
XMM-Newton} data). Our data from the deep {\it Chandra} observations show 
evidence of Fe K line emission from the highly-ionized H-like Fe atoms, which 
confirms the results reported with the {\it XMM-Newton} data. The detected H-like 
Fe K line feature ($E$ $\sim$ 6.96 keV) requires an extremely high temperature 
plasma of $kT$ $\sim$ 5 keV in addition to the $kT$ $\sim$ 1 keV component, 
which also confirms the results obtained with the {\it XMM-Newton}. We thus 
attribute the discrepancies between M02 and S04 on the H-like Fe line features 
to the poor photon statistics of the previous {\it Chandra} data. The hard
component plasma temperature is, however, unusually high for an SNR, and the 
interpretation of its origin is difficult. The peculiar interstellar 
environments in the Galactic center, such as unusually high magnetic fields, 
might be responsible for this high temperature. For instance, Muno et al. (2004a) 
reported the existence of very high temperature ($kT$ $\sim$ 8 keV) thermal plasma 
prevailing in the Galactic center regions. 

On the other hand, we detect no significant evidence for the neutral Fe line 
emission from Sgr A East, in contrast to the results from the {\it XMM-Newton} 
observation. It is notable that the sky position of Sgr A East is on the 
northeastern side of Sgr A*, where the background 6.4 keV neutral Fe line emission 
(and other emission line features) of the Galactic center region prevails 
\citep{park04a}. The reported detection of the neutral Fe line with the {\it 
XMM-Newton} might have thus resulted from the inclusion of background diffuse 
emission and point sources in the spectrum of Sgr A East, neither of which could 
be properly accounted for, because of the poorer angular resolution of {\it 
XMM-Newton}. It is important to {\it avoid} confusion by the complex background 
line emission in the Galactic center for the spectral analysis of Sgr A East. 
The deep exposure and the superb angular resolution of our {\it Chandra} data 
allowed us to use small regions for the spectral analysis. For instance, our 
``center'' region is an order of magnitude smaller in the source area than the 
central region used by S04, while achieving good photon statistics. We also 
utilized background regions which thoroughly represent {\it all} emission line 
features around Sgr A East. The detected point sources were not removed in the 
{\it XMM-Newton} data analysis (S04). Because faint point sources in the 
Galactic center regions are typically neutral Fe line sources \citep{muno04b}, 
the contribution from the unremoved point sources in the {\it XMM-Newton} data 
may also be responsible for the reported 6.4 keV line there (S04). 
In order to test these technical differences between S04 and the current work, 
we extracted the spectrum from a relatively large-size region 
($\sim$28$^{\prime\prime}$ radius) around the center of Sgr A East without 
removing the detected point sources. We used the {\it same} background region 
as used in S04. We then made an $\sim$4$\sigma$ {\it detection} of the 6.4 keV 
line. The best-fit photon flux from the neutral Fe line was $\sim$2.5 $\times$ 
10$^{-6}$ photons cm$^{-2}$ s$^{-1}$ which was consistent with that reported 
by S04. We thus conclude that the previously reported 6.4 keV line emission 
from Sgr A East with the {\it XMM-Newton} data (S04) was residual contamination 
from the Galactic center background emission.

\section{\label{sec:summary} SUMMARY AND CONCLUSIONS}

We presented the results from imaging and spectral analyses of the SNR, Sgr A 
East in the Galactic center, using deep {\it Chandra} observations.
We confirm the central concentration of Fe-rich stellar ejecta in Sgr A 
East. The X-ray spectrum from the central regions of the SNR shows X-ray line
emission from highly ionized He-like and H-like Fe. These Fe line features
require multiple temperature thermal plasma with $kT$ $\sim$ 1 keV and 5 keV
in order to properly describe the observed spectrum. The strong Fe lines
imply overabundant Fe in the center of the SNR. On the other hand, the soft X-ray 
emission extending to the northern side of the SNR can be described with a single
temperature ($kT$ $\sim$ 1.3 keV) thermal plasma with solar abundances. This
northern ``plume'' is thus likely X-ray emission from the shocked ISM rather
than metal-rich SN ejecta. 

A hard {\it pointlike} source (the so-called {\it cannonball}) detected at the 
northern edge of the plume, designated as CXOGC J174545.5$-$285829 in the 
Galactic center source catalog, shows remarkably unusual X-ray characteristics. 
The morphological, spectral, and temporal properties of this source are 
suggestive of an identification as a high-velocity neutron star. The estimated 
Fe ejecta mass of Sgr A East is consistent with a Type II SN for the origin 
of Sgr A East. Based on the suggested Type II origin for Sgr A East, the likely 
identification of the {\it cannonball} as a high-velocity neutron star, and the 
proximity between the {\it cannonball} and Sgr A East, we propose that these 
objects comprise an SNR-NS association in the Galactic center. We note, however, 
that follow-up pulsar searches with high time-resolution X-ray and/or radio 
observations at the position of the {\it cannonball} will be needed in order 
to make a conclusive identification of a neutron star for this object.

\acknowledgments
This work has been supported in parts by NASA contract 
NAS8-39073, NAS8-01128 for the {\it Chandra X-Ray Observatory}.
MPM is supported by the Hubble Fellowship program through grant number
HS-HF-01164.01-A.

\clearpage

\begin{deluxetable}{ccc}
\footnotesize
\tablecaption{Chandra/ACIS Observation log
\label{tbl:tab1}}
\tablewidth{0pt}
\tablehead{\colhead{ObsID} & \colhead{Date} & \colhead{Exposure (ks)}}
\startdata
0242 & 1999-09-21 & 41 \\
1561a & 2000-10-26 & 36 \\
1561b & 2001-07-14 & 14 \\
2943 & 2002-05-22 & 35 \\
2951 & 2002-02-19 & 12 \\
2952 & 2002-03-23 & 12 \\
2953 & 2002-04-19 & 12 \\
2954 & 2002-05-07 & 12 \\
3392 & 2002-05-25 & 167 \\
3393 & 2002-05-28 & 158 \\
3663 & 2002-05-24 & 38 \\
3665 & 2002-06-03 & 90 \\
\enddata
\end{deluxetable}

\begin{deluxetable}{cccc}
\footnotesize
\tablecaption{Energy Bands used for Generating the Iron Equivalent Width
Image.
\label{tbl:tab2}}
\tablewidth{0pt}
\tablehead{ \colhead{Elements} & \colhead{Line} &
\colhead{Low\tablenotemark{a}} & \colhead{High\tablenotemark{a}} \\
 & \colhead{(eV)} & \colhead{(eV)} & \colhead{(eV)} }
\startdata
Fe (He$\alpha$) & 6510 $-$ 6690 & 5000 $-$ 5730 & 7470 $-$ 8500 \\
\enddata
\tablenotetext{a}{The high and low energy bands around the
selected line energies used to estimate the underlying continuum.}

\end{deluxetable}

\begin{deluxetable}{ccccccc}
\footnotesize
\tablecaption{Results of Spectral Fittings of Sgr A East\tablenotemark{a}
\label{tbl:tab3}}
\tablewidth{0pt}
\tablehead{\colhead{Region} & \colhead{$N_H$} & \colhead{$kT_{soft}$} &
\colhead{$kT_{hard}$} & \colhead{$EM_{soft}$} & \colhead{$EM_{hard}$} & 
\colhead{$\chi^{2}$/$\nu$} \\
     & \colhead{(10$^{22}$ cm$^{-2}$)} & \colhead{(keV)} & \colhead{(keV)} 
& \colhead{(10$^{57}$ cm$^{-3}$)} & \colhead{(10$^{57}$ cm$^{-3}$)} & } 
\startdata
Center & 18.9$^{+2.3}_{-1.8}$ & 1.05$^{+0.19}_{-0.13}$ & 5.28$^{+0.66}_{-0.57}$ 
& 7.1$^{+5.5}_{-2.8}$ & 0.3$^{+0.1}_{-0.1}$ & 167.9/163 \\
Plume & 14.7$^{+1.5}_{-1.9}$ & 1.32$^{+0.24}_{-0.11}$ & - & 
1.8$^{+0.6}_{-0.6}$ & - & 100.5/83 \\
North & 13.0$^{+1.0}_{-1.0}$ & 1.00$^{+0.15}_{-0.10}$ & 10.9$^{+19.1}_{-4.3}$ &
3.7$^{+1.9}_{-1.3}$ & 0.05$^{+0.04}_{-0.02}$ & 139.7/110 \\
\enddata
\tablenotetext{a}{The errors are 2$\sigma$ uncertainties.} 
\end{deluxetable}

\begin{deluxetable}{ccccc}
\footnotesize
\tablecaption{Best-Fit Metal Abundances of Sgr A East\tablenotemark{a,b}
\label{tbl:tab4}}
\tablewidth{0pt}
\tablehead{\colhead{Region} & \colhead{S} & \colhead{Ar} & \colhead{Ca} &
\colhead{Fe}} 
\startdata
Center & 0.69$^{+0.74}_{-0.56}$ & 1.76$^{+0.80}_{-0.72}$ & 
1.37$^{+0.80}_{-0.76}$ & 5.81$^{+1.65}_{-1.11}$ \\
Plume & 1.91$^{+0.98}_{-0.69}$ & 1.20$^{+0.65}_{-0.58}$ &
2.50$^{+0.77}_{-0.71}$ & 0.96$^{+0.38}_{-0.27}$ \\
North & 1.00$^{+0.31}_{-0.28}$ & 1.09$^{+0.44}_{-0.44}$ &
1.55$^{+0.63}_{-0.59}$ & 2.49$^{+1.36}_{-0.85}$ \\
\enddata
\tablenotetext{a}{Abundances are with respect to solar.}\\
\tablenotetext{b}{The errors are 2$\sigma$ uncertainties.}
\end{deluxetable}

\begin{deluxetable}{ccccc}
\footnotesize
\tablecaption{Results of Spectral Fittings of The {\it Cannonball}
\label{tbl:tab5}}
\tablewidth{0pt}
\tablehead{\colhead{$\Gamma$\tablenotemark{a}} & 
\colhead{$N_H$\tablenotemark{a}} & 
\colhead{$f_{X (2-10 keV)}$\tablenotemark{a}} & \colhead{$L_{X (2-10 keV)}$} & 
\colhead{$\chi^{2}$/$\nu$} \\
& \colhead{(10$^{22}$ cm$^{-2}$)} & 10$^{-13}$ ergs cm$^{-2}$ s$^{-1}$ & 
10$^{33}$ ergs s$^{-1}$ &}
\startdata
1.59$^{+0.58}_{-0.39}$ & 16.5$^{+3.3}_{-3.2}$ & 1.9$^{+3.6}_{-1.2}$ & 
3.1 & 37.0/46 \\
\enddata
\tablenotetext{a}{The errors are 2$\sigma$ uncertainties.}
\end{deluxetable}

\clearpage

\end{document}